\documentclass[modern]{aastex61}

\usepackage{graphicx}
\usepackage{amsmath}
\usepackage{amssymb}
\usepackage{enumitem}

\newcommand\mybar{\kern1pt\rule[-\dp\strutbox]{.8pt}{\baselineskip}\kern1pt}

\setlist[itemize]{noitemsep, topsep=0pt, leftmargin=*}

\shorttitle{Expansion or Contraction of Galaxies}
\shortauthors{Loeb}

\begin{document}

\title{Measuring the Expansion or Contraction of Galaxies}

\author{Abraham Loeb}
\affil{Astronomy Department, Harvard University, 60 Garden St., Cambridge, MA 02138, USA}

\begin{abstract}
Galaxies lose mass as a result of their luminosity or gaseous
outflows.  I calculate the resulting radial migration of stars
outwards and show that it could potentially be measured with high resolution
spectrographs on the next generation of large telescopes. Substantial
accretion of matter in dense cosmic environments could trigger inward
stellar migration that would be even more easily measurable.

\end{abstract}

\section*{}
The bolometric luminosity $L$ of galaxies leads to an inevitable mass
loss rate,
\begin{equation}
{\dot M}=-{L\over c^2}= 3.5\times 10^6 \left({\rm M_\odot\over Gyr}\right)\left({L\over L_{\rm MW}}\right),
\label{mdot}
\end{equation}
where the luminosity normalization on the right-hand-side is the
intergal over the spectral energy distribution of the Milky-Way
galaxy, $L_{\rm MW}\approx 5\times 10^{10} L_\odot$ \citep{2021MNRAS.508.4459F}.

Stars on circular orbits with a velocity $v_{\rm c}$ and orbital
radius $r$ in the outskirt of a galaxy, like the Sun in the Milky-Way,
will drift outwards as a result of this mass loss. Since the fraction
of mass lost per stellar orbital time $(2\pi r/v_{\rm c})$ is much
smaller than unity, the orbital angular momentum per unit mass
$j=v_{\rm c}r$ is a conserved adiabatic invariant. Therefore, the
orbital radius of a circular orbit maintains the relation,
$r=j^2/GM_{\rm tot}$, where $M_{\rm tot}$ is the total mass (including
dark matter) enclosed within a radius $r$. The mass loss rate,
${\dot M}$, results in radial migration outwards at a speed,
\begin{equation}
v_r\equiv {dr\over dt}= -\left({{\dot M}\over M_{\rm tot}}\right)r .
\label{vr}
\end{equation}

Combinining equation (\ref{vr}) with the dynamical relation for a flat
rotation curve,
\begin{equation}
M_{\rm tot}={v_{\rm c}^2 r\over G},
\label{vcirc}
\end{equation} 
and the Tully-Fisher relation for nearby disk galaxies \citep{2021AJ....162..202M}, 
\begin{equation}
\left({L\over L_{\rm MW}}\right)=\left({v_{\rm c}\over v_{\rm MW}}\right)^4,
\label{tf}
\end{equation}
where $v_{\rm MW}\approx 236~{\rm km~s^{-1}}$
\citep{2019ApJ...885..131R}, yields,
\begin{equation}
v_r=27~{\rm cm~s^{-1}}\left({L\over L_{\rm MW}}\right)^{1/2} .
\label{finalvr}
\end{equation}

Integrated over a time interval $t$, this drift leads to a radial displacement of
\begin{equation}
\Delta r=2.7~{\rm pc}\left({t\over 10 {\rm Gyr}}\right)\left({L\over L_{\rm
    MW}}\right)^{1/2} .
\label{finalvr}
\end{equation}

The most luminous disk galaxies in the local universe
\citep{2019ApJS..243...14O} have luminosities as high as $\sim
20L_{\rm MW}$, yielding radial expansion speeds of up to about a meter
per second. Such speeds will be detectable by high resolution
spectrographs on the next generation of ground-based extremely large 
telescopes \citep{2021Msngr.182...27M,2018arXiv180905804C}.

Higher expansion speeds are expected to be triggered during bright
quasar episodes, when the supermassive black hole luminosity could
exceed the total stellar luminosity of the host galaxy by two orders
of magnitude, leading to stellar migration outwards of up to $\sim 0.1
{\rm km~s^{-1}}$ during the episodic quasar lifetime of typically $\sim
0.1~{\rm Gyr}$ \citep{2021MNRAS.505..649K}. Outflows of gas could
increase the outward migration by another order of magnitude.

For quiescent galaxies, inward migration is expected to result
from the accretion of intergalactic matter or galaxy mergers. The
radial migration of stars would be easily detectable in dense
environments, like galaxy groups or large-scale filaments, where the
inward velocity could reach values as high as,
\begin{equation}
v_r= -0.8~{\rm km~s^{-1}}\left({{\dot M}\over 10{\rm M_{\odot}~yr^{-1}}}\right)\left({L\over L_{\rm MW}}\right)^{1/2}.
\label{vr}
\end{equation}

\bigskip
\bigskip

\section*{Acknowledgements}

This work was supported in part by Harvard's {\it Black Hole
  Initiative}, which is funded by grants from JFT and GBMF.

\bibliographystyle{aasjournal}
\bibliography{exp}
\label{lastpage}
\end{document}